\documentclass[aps,prl,twocolumn,floatfix,superscriptaddress]{revtex4-2}
\usepackage{graphicx,amsmath,amsfonts,mathrsfs,bbold,dsfont}
\begin{document}

\title{Altermagnetism and Weak Ferromagnetism}

\author{I.~V.~Solovyev}
\email{SOLOVYEV.Igor@nims.go.jp}
\affiliation{Research Center for Materials Nanoarchitectonics (MANA), National Institute for Materials Science (NIMS), 1-1 Namiki, Tsukuba, Ibaraki 305-0044, Japan}
\author{S.~A.~Nikolaev}
\affiliation{Department of Materials Engineering Science, The University of Osaka, Toyonaka 560-8531, Japan}
\author{A.~Tanaka}
\affiliation{Research Center for Materials Nanoarchitectonics (MANA), National Institute for Materials Science (NIMS), 1-1 Namiki, Tsukuba, Ibaraki 305-0044, Japan}

\date{\today}

\date{\today}
\begin{abstract}
Using a realistic model relevant to La$_2$CuO$_4$ and other altermagnetic perovskite oxides, we study interrelations between weak ferromagnetism (WF), anomalous Hall effect (AHE), and net orbital magnetization (OM). All of them can be linked to the form of Dzyaloshinskii-Moriya (DM) interactions. Nevertheless, while spin WF is induced by the DM vector components having the same sign in all equivalent bonds, AHE and OM are related to sign-alternating components, which do not contribute to any canting of spins. The microscopic model remains invariant under the symmetry operation $\{ \mathcal{S}|{\bf t} \}$, combining the shift ${\bf t}$ of antiferromagnetically coupled sublattices to each other with the spin flip $\mathcal{S}$. Thus, the band structure remains spin-degenerate, but the time-reversal symmetry is broken, providing a possibility to realize AHE in antiferromagnetic substances. The altermagnetic splitting of bands, breaking the $\{ \mathcal{S}|{\bf t}\}$ symmetry, does not play a major role in the problem. More important is the orthorhombic strain, responsible for finite values of AHE and OM.
\end{abstract}

\maketitle

\par \emph{Introduction}. Altermagnetism is regarded as a new phase of matter, where a nearly antiferromagnetic (AFM) alignment of spins coexists with the spin splitting of bands and robust time-reversal symmetry ($\mathcal{T}$) breaking, which are typical for ferromagnetic (FM) systems~\cite{SmejkalPRX1,SmejkalPRX2,LingBai,Naka_Spintronics}. Nevertheless, this new classification raises new questions, especially on how it fits into our previous knowledge on AFM materials. It is certainly true that the lifting of Kramers' degeneracy of AFM bands, which was predicted largely due to development of density-functional theory (DFT) calculations, is a new aspect of the problem~\cite{Noda,Okugawa,HayamiJPSJ,Naka,SmejkalSA,NakaOrganic}. On the other hand, the possibility of breaking $\mathcal{T}$ in certain classes of AFM systems has been known for decades. As was pointed out by Dzyaloshinskii~\cite{Dzyaloshinskii1991}, using phenomenological symmetry arguments, the materials whose magnetic unit cell coincides with the crystallographic one present a special type of antiferromagnetism, giving rise to such phenomena as weak ferromagnetism (WF)~\cite{Dzyaloshinskii_weakF}, piezomagnetism~\cite{DzyaloshinskiiPM}, and magnetoelectricity~\cite{DzyaloshinskiiME}. A very detailed classification was given by Turov~\cite{TurovBook,TurovUFN}, who argued that there are two major classes of unconventional antiferromagnets, depending on whether the spatial inversion $\mathcal{I}$ enters the magnetic group in combination with $\mathcal{T}$ or alone. The first scenario corresponds to magnetoelectricity, while the second one, which encompasses WF and piezomagnetism, has clear similarity to what is now called ``altermagnetism''. Canonically, WF refers to net spin magnetic moments in otherwise AFM substances, while altermagnetism emerged from the analysis of the anomalous Hall effect (AHE)~\cite{SmejkalSA,NakaOrganic}. However, from the phenomenological point of view, these two effects are identical to each other: both manifest that $\mathcal{T}$ is macroscopically broken and the existence of AHE automatically implies the existence of WF, no matter how ``weak'' it is. Actually, in his monograph~\cite{TurovBook}, Turov considered not only WF, but also AHE and many other phenomena expected in weak ferromagnets. Particularly, already in 1962, he and Shavrov predicted that AHE can be induced by AFM order~\cite{TurovShavrov}. Apparently, this is the most one can say within phenomenological theories, which do not provide any information about the magnitude of the effect or its microscopic origin.

\par Despite the fact that AHE is regarded to be inherent to altermagnetic systems, the aspect of spin splitting in this phenomenon remains to be obscured, and so does the general relation between WF and altermagnetism. The microscopic theory of WF is basically the theory of Dzyaloshinskii-Moriya (DM) interactions that was proposed by Moriya for Mott insulators~\cite{Moriya_weakF} and extended for magnetic systems~\cite{Katsnelson,Kikuchi,review2024}. Depending on the symmetry, DM interactions can have several components: those that have the same sign in equivalent bonds support WF, while the ones with alternating signs do not contribute to the FM moment or any canting of spins. A notable example is the DM interactions between two magnetic sublattices in CrO$_2$, a sister compound of altermagnetic RuO$_2$: these interactions are relatively strong but alternate among eight neighbouring bonds~\cite{review2024}. Regarding the role played by the sign-alternating DM interactions, we will argue that they do bear direct and crucial implications for AHE and net orbital magnetization ${\cal M}$, thus clarifying the fundamental difference between AHE and WF from the microscopic point of view. On the other hand, while the lifting of Kramers spin degeneracy in the non-relativistic limit is regarded to be the central manifestation of breaking $\mathcal{T}$ in altermagnets~\cite{SmejkalPRX1,SmejkalPRX2,LingBai}, we will show that: (i) the bands can remain spin-degenerate even when $\mathcal{T}$ is broken and (ii) the spin splitting does not play an essential role in the emergence of AHE and ${\cal M}$.

\par Finally, in 1997, long before the modern era of altermagnetism, one of us proposed that weak ferromagnets La$M$O$_3$ ($M=$ Cr, Mn, and Fe) can exhibit an appreciable magneto-optical effect (the ac analog of AHE)~\cite{PRB1997}. Rather than relating to the WF itself, this phenomenon better correlated with the behavior of orbital magnetic moments~\cite{PRB1997}. Using the modern theory of orbital magnetization~\cite{Thonhauser,Shi}, we will argue that the conclusion was essentially correct as AHE and ${\cal M}$ have a similar microscopic origin, and ${\cal M}$ can be regarded as a proper order parameter for altermagnets. 

\par \emph{Lattice symmetry}. To be specific, we keep in mind the $B$O$_2$ layer of orthorhombic perovskites $AB$O$_3$ with the space group $Pbnm$ or layered perovskites $A_2B$O$_4$ with the space group $Bmab$. The characteristic examples are LaFeO$_3$ and La$_2$CuO$_4$, forming AFM order in the layer plane ($xy$). There are two magnetic sublattices, centered at the origin $(0,0)$ and ${\bf t}=(\frac{1}{2},\frac{1}{2})$. The $Pbnm$ group has four symmetry operations transforming the plane $xy$ to itself: the unity; $\mathcal{I}$; the twofold rotation about $x$ combined with the shift by $\bf{t}$, $\{ \mathcal{C}_{2x}|{\bf t}\}$; and the mirror reflection of $x$, also combined with the shift, $\{ m_{x}|{\bf t}\}$. In $Bmab$, these symmetry operations are combined with the mirror reflection $m_{y}$~\cite{footnote2}. $\{ \mathcal{C}_{2x}|{\bf t}\}$ and $\{ m_{x}|{\bf t}\}$ transform two magnetic sublattices to each other, while $\mathcal{I}$ and $m_{y}$ transforms each sublattice to itself. 

\par \emph{Model}. The microscopic model for the $B$O$_2$ layer includes the following ingredients (see Fig.~\ref{fig:model}): 
\noindent
\begin{figure}[b]
\begin{center}
\includegraphics[width=8.6cm]{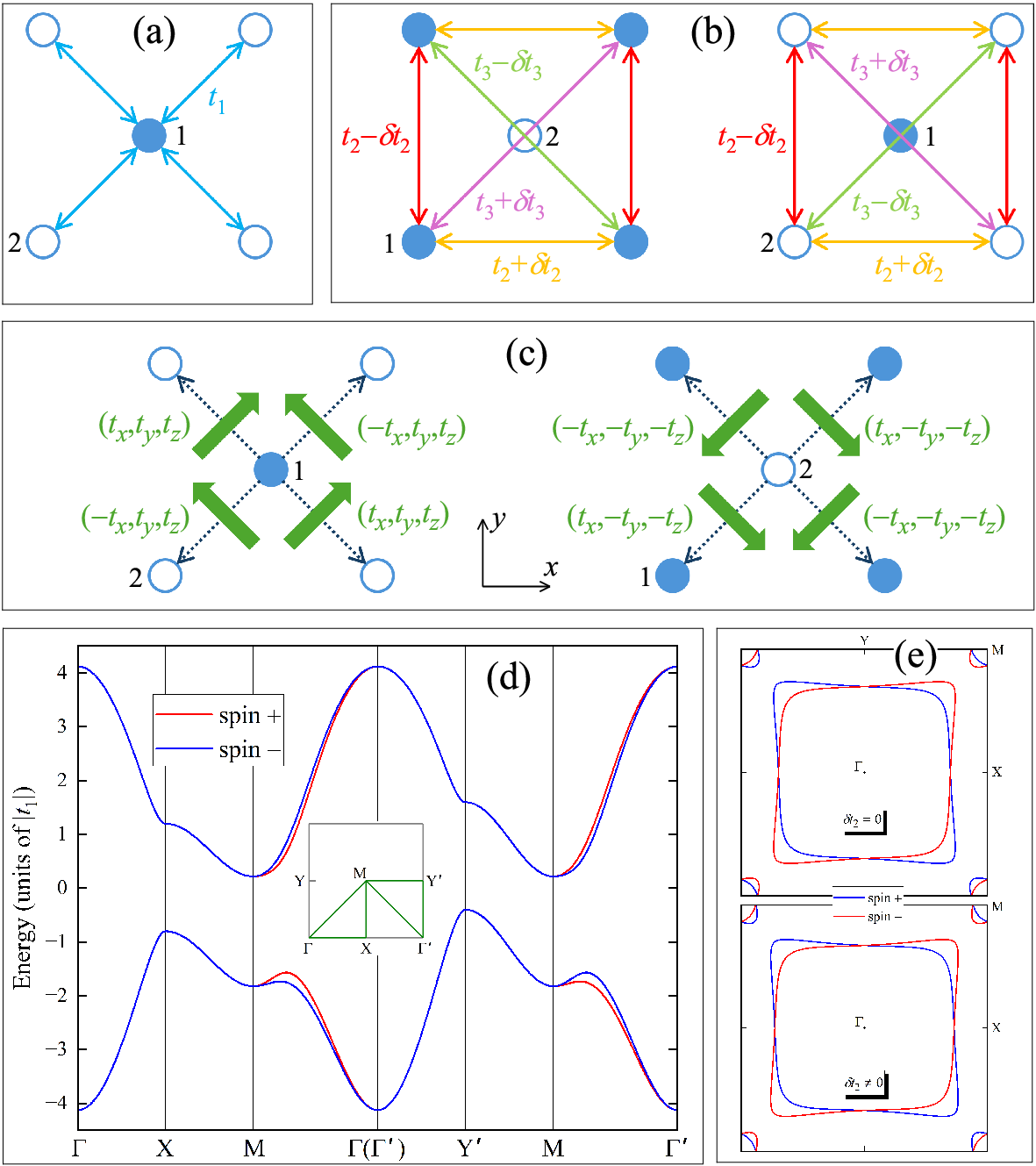} 
\end{center}
\caption{Parameters and basic electronic structure: Hopping parameters between (a) first nearest neighbors and (b) second and third nearest neighbors; (c) Parameters of spin-orbit interaction around two magnetic sites (denotes as $1$ and $2$). The directions of the bonds are shown by dotted arrows; (d) Example of band structure. The inset shows high-symmetry points of the Brillouin zone; and (e) Corresponding Fermi surface for $n_{\rm el}=1$ with and without orthorhombic strain $\delta t_{2}$.}
\label{fig:model}
\end{figure}
\noindent(i) The hoppings between first, second, and third nearest neighbors ($t_{1}$, $t_{2}$, and $t_{3}$, respectively). (ii) The orthorhombic strain of the second nearest hoppings, $\delta t_{2}$, having the same form in both magnetic sublattices and making the directions $x$ and $y$ inequivalent. As we will see, this is a very important parameter, which is responsible for finite values of AHE. (iii) Deformation of the third nearest hoppings, $\delta t_{3}$, alternating between the sublattices $1$ and $2$, that results in the altermagnetic splitting of bands. (iv) SOC in noncentrosymmetric nearest-neighbor bonds, which has the form of spin-dependent hoppings, $\hat{\cal H}^{\rm so}_{ij} = i\boldsymbol{t}_{ij} \cdot \hat{\boldsymbol{\sigma}}$, where $\hat{\boldsymbol{\sigma}} = (\hat{\sigma}_{x},\hat{\sigma}_{y},\hat{\sigma}_{z})$ is the vector of spin Pauli matrices. The DM interaction, $\boldsymbol{D}_{ij}$, is simply proportional to $\boldsymbol{t}_{ij}$ and this universal property holds in insulating~\cite{Shekhtman} as well as metallic~\cite{Katsnelson,Kikuchi,review2024} regimes. In orthorhombic systems, $\boldsymbol{D}_{ij} \sim \boldsymbol{t}_{ij}$ have the following form around each magnetic site~\cite{PRL1996}: $\boldsymbol{t}_{ij} = (\pm t_{x},t_{y},t_{z})$, where $y$ and $z$ components are the same in all the bonds, while the sign of $x$ component alternates as shown in Fig.~\ref{fig:model}(c). Thus, if AFM spins are aligned along $x$, $t_{y}$ and $t_{z}$ are responsible for the WF along, respectively, $z$ and $y$~\cite{Yamaguchi}, while $t_{x}$ has no effect on the spin texture. The $Bmab$ symmetry imposes additional constraints: $\delta t_{3}=0$ and $t_{z}=0$~\cite{footnote3}. Then, $\hat{\cal H}^{\rm so}_{ij}$ can be eliminated by the ${\rm SU(2)}$ rotations of the spins, $\hat{U}_{S} = e^{-i \varphi \boldsymbol{n} \cdot \hat{\boldsymbol{\sigma}}}$, in the sublattice $2$ with $\boldsymbol{n}=\frac{\boldsymbol{t}_{ij}}{|\boldsymbol{t}_{ij}|}$, $\varphi = 2 \arctan \frac{|\boldsymbol{t}_{ij}|}{t_{1}}$, which leads to the redefinition $t_{1} \to t_{1}\sqrt{1+\left(\boldsymbol{t}_{ij}/t_{1}\right)^{2}}$~\cite{Shekhtman,Kaplan}. However, this transformation depends on the bond and, therefore, cannot be performed simultaneously for all the bonds. Nevertheless, we can still use it to eliminate $t_{y}$ and $t_{z}$, but to keep $t_{x}$ and $\hat{\cal H}^{\rm so}_{ij} = \pm it_{x} \hat{\sigma}_{x}$. The corresponding transformation is given by $\hat{U}_{S}$ with $t_{x}=0$. Therefore, the parameters $t_{y}$ and $t_{z}$, which are responsible for spin WF, do not play any role in AHE and ${\cal M}$. Finally, we add the N\'eel field $\pm B \hat{\sigma}_{x}$ along $x$, which yields FM moment along $z$~\cite{Yamaguchi}. Then, after global rotation of spins, transforming $\hat{\sigma}_{x}$ to $\hat{\sigma}_{z}$, we will have the following $4$$\times$$4$ Hamiltonian~\cite{Roig}, which is diagonal with respect to spins $\sigma = \pm$:
\noindent
\begin{equation}
\hat{\cal H}_{\boldsymbol{k}} = h^{\phantom{0}}_{\boldsymbol{k}} - \delta h^{3}_{\boldsymbol{k}} \hat{\tau}_{z} + h^{1}_{\boldsymbol{k}} \hat{\tau}_{x}  -B \hat{\tau}_{z} \hat{\sigma}_{z} - h^{\rm so}_{\boldsymbol{k}}\hat{\tau}_{y} \hat{\sigma}_{z} ,
\label{eq:Hk}
\end{equation}
\noindent where two AFM sublattices are described in terms of Pauli matrices $\hat{\boldsymbol{\tau}} = (\hat{\tau}_{x},\hat{\tau}_{y},\hat{\tau}_{z})$~\cite{Roig}, $h^{\phantom{0}}_{\boldsymbol{k}} = h^{2}_{\boldsymbol{k}} + \delta h^{2}_{\boldsymbol{k}} + h^{3}_{\boldsymbol{k}}$, $h^{2}_{\boldsymbol{k}} = 2t_{2} (\cos k_{x} + \cos k_{y})$, $\delta h^{2}_{\boldsymbol{k}} = \delta t_{2}(\cos k_{x} - \cos k_{y})$, $h^{3}_{\boldsymbol{k}} = 4t_{3} \cos k_{x} \cos k_{y}$, $\delta h^{3}_{\boldsymbol{k}} = 4 \delta t_{3} \sin k_{x} \sin k_{y}$, $h^{1}_{\boldsymbol{k}} = 4t_{1} \cos \frac{k_{x}}{2} \cos \frac{k_{y}}{2}$, and $h^{\rm so}_{\boldsymbol{k}} = 4t_{x} \sin \frac{k_{x}}{2} \sin \frac{k_{y}}{2}$. 

\par The model parameters can be derived from DFT~\cite{WannierRevModPhys,wannier,rpa2,respack2}, as we will do below for La$_2$CuO$_4$. Alternatively, one can take the exchange interactions, $J_{k}$, and evaluate the parameters from the superexchange theory as $|t_{k}/t_{1}| = \sqrt{J_{k}/J_{1}} $ and $2 \boldsymbol{t}_{ij}/t_{1} = \boldsymbol{D}_{ij}/J_{1}$~\cite{Shekhtman}. $\delta t_{2}$ and $\delta t_{3}$ can be exceptionally large in orthorhombic manganites, to facilitate the formation of spin-spiral multiferroic phases~\cite{review2024}. Then, a reasonable choice is (in units of $t_{1}<0$): $t_{3} \sim -t_{2} \sim 0.1$ and $ \delta t_{3} \sim t_{x} \sim -\delta t_{2} \sim 0.05$~\cite{review2024}, which will be used unless it is specified otherwise. $|B|=1$ is sufficient to open the band gap. Examples of such band structure, $\varepsilon^{\sigma}_{\boldsymbol{k},\nu} = h^{\phantom{0}}_{\boldsymbol{k}} + \nu \sqrt{(\sigma B+\delta h^{3}_{\boldsymbol{k}})^{2} + ( h^{1}_{\boldsymbol{k}})^{2} + (h^{\rm so}_{\boldsymbol{k}})^{2} }$ ($\nu=\pm$ being the band index), and Fermi surface are shown in Fig.~\ref{fig:model}(d),(e). As expected, $\delta t_{3}$ splits the bands, while $\delta t_{2}$ deforms the Fermi surface.

\par \emph{Hidden symmetries}. A very interesting situation occurs when $\delta t_{3}=0$ (i.e., \emph{without altermagnetic splitting of bands}). Since $\mathcal{T} \hat{\sigma}_{z}=-\hat{\sigma}_{z}$, the N\'eel field $B \hat{\tau}_{z} \hat{\sigma}_{z}$ breaks $\mathcal{T}$, but remains invariant under $\{ \mathcal{T}|{\bf t}\}$, where $\mathcal{T}$ is combined with the lattice shift of site $1$ to site $2$ (and vice versa), which additionally changes $\hat{\tau}_{z}$ to $-\hat{\tau}_{z}$. Then, since $B \hat{\tau}_{z} \hat{\sigma}_{z}$ is real and $\mathcal{T}=\mathcal{S}K$, where $K$ is the complex conjugation and $\mathcal{S} = i\hat{\sigma}_{y}$ flips $\sigma$ to $-\sigma$, $\mathcal{T}$ can be replaced by $\mathcal{S}$, so that the N\'eel field is also invariant under $\{ \mathcal{S}|{\bf t}\} \equiv i \hat{\sigma}_{y} \hat{\tau}_{x}$. On the other hand, the SO interaction, $h^{\rm so}_{\boldsymbol{k}}\hat{\tau}_{y} \hat{\sigma}_{z}$, is invariant under $\mathcal{T}$, but changes its sign when it is combined with the lattice shift, $\{ \mathcal{T}|{\bf t}\}$. However, since $\hat{\tau}_{y}$ is complex, this sign change does not occur if one uses $\{ \mathcal{S}|{\bf t}\}$ instead of $\{ \mathcal{T}|{\bf t}\}$. Therefore, the SO interaction is invariant under $\{ \mathcal{S}|{\bf t}\}$, so as the full Hamiltonian (\ref{eq:Hk}). This means that eigenstates with $\sigma = \pm$ differ only by a phase factor. The latter guarantees that (i) the $\sigma = \pm$ bands are degenerate and (ii) the contributions of these bands to $\sigma_{xy}$ are equal to each other and, instead of the partial cancellation, which would occur in ferromagnets, we have an \emph{addition} of such contributions. 

\par \emph{Berry curvature and AHE}. $\sigma_{xy}$ is given by the Brillouin zone (BZ) integral of the Berry curvatures, $\Omega^{\sigma}_{\boldsymbol{k}}$~\cite{Fang,Yao} (in atomic units): 
\noindent
\begin{displaymath}
\sigma_{xy} =  - \int_{\rm BZ} \frac{d \boldsymbol{k}}{(2 \pi)^{2}} \sum_{\sigma = \pm} f^{\sigma}_{\boldsymbol{k}} \Omega^{\sigma}_{\boldsymbol{k}} , 
\end{displaymath}
\noindent where $f^{\sigma}_{\boldsymbol{k}}$ is Fermi-Dirac distribution function and $\Omega^{\sigma}(\boldsymbol{k}) = -2 {\rm Im} \left\langle \partial_{k_{x}} u^{\sigma}_{\boldsymbol{k}} | \partial_{k_{y}} u^{\sigma}_{\boldsymbol{k}} \right\rangle$. Without loss of generality, one can consider the case where $\nu = -$ bands are partially occupied by $n_{\rm el}$ electrons while $\nu = +$ bands are empty. Searching eigenvectors for $\nu = -$ in the form: 
\noindent
\begin{displaymath}
| u^{\sigma}_{\boldsymbol{k}} \rangle = \left(
\begin{array}{c}
\cos \theta^{\sigma}_{\boldsymbol{k}} e^{i \phi^{\sigma}_{\boldsymbol{k}}} \\
\sin \theta^{\sigma}_{\boldsymbol{k}}
\end{array}
\right), 
\end{displaymath}
\noindent it is straightforward to find~\cite{SM}: 
\noindent
\begin{displaymath}
\theta^{\sigma}_{\boldsymbol{k}}  = - \frac{1}{2} \arctan \frac{ \sqrt{( h^{1}_{\boldsymbol{k}})^{2} + (h^{\rm so}_{\boldsymbol{k}})^{2}}} {\sigma B + \delta h^{0}_{\boldsymbol{k}}} 
\end{displaymath}
\noindent ($0$$\le$$\theta^{\sigma}_{\boldsymbol{k}}$$<$$\pi$) and $\phi^{\sigma}_{\boldsymbol{k}} = \sigma \arctan \left( h^{\rm so}_{\boldsymbol{k}}/h^{1}_{\boldsymbol{k}} \right)$ ($0$$\le$$\phi^{\sigma}_{\boldsymbol{k}}$$<$$2\pi$), which are, respectively, even and odd in $t_{x}$ (SOC). Therefore, the Berry curvature, which can be expressed as the cross product, $\Omega^{\sigma}(\boldsymbol{k}) = \sin 2 \theta^{\sigma}_{\boldsymbol{k}} \, \left[\partial^{\phantom{\sigma}}_{\boldsymbol{k}} \theta^{\sigma}_{\boldsymbol{k}} \times \partial^{\phantom{\sigma}}_{\boldsymbol{k}} \phi^{\sigma}_{\boldsymbol{k}} \right]_{z}$, is odd in $t_{x}$. Since $\sin 2 \theta^{\sigma}_{\boldsymbol{k}} \, \partial^{\phantom{\sigma}}_{\boldsymbol{k}} \theta^{\sigma}_{\boldsymbol{k}} = -\frac{1}{2} \partial^{\phantom{\sigma}}_{\boldsymbol{k}} \left( \cos 2 \theta^{\sigma}_{\boldsymbol{k}} \right)$, $\sigma_{xy}$ can be reformulated in terms of group velocities, $\partial^{\phantom{\sigma}}_{\boldsymbol{k}} \varepsilon^{\sigma}_{\boldsymbol{k}}$, at the Fermi surface~\cite{SM}: 
\noindent
\noindent
\begin{displaymath}
\sigma_{xy} =  -\frac{1}{2} \int_{\rm BZ} \frac{d \boldsymbol{k}}{(2 \pi)^{2}}  \sum_{\sigma =\pm} \frac{\partial f^{\sigma}_{\boldsymbol{k}}}{\partial \varepsilon^{\sigma}_{\boldsymbol{k}}}  \cos 2 \theta^{\sigma}_{\boldsymbol{k}} \left[  \partial^{\phantom{\sigma}}_{\boldsymbol{k}} \varepsilon^{\sigma}_{\boldsymbol{k}} \times \partial^{\phantom{\sigma}}_{\boldsymbol{k}} \phi^{\sigma}_{\boldsymbol{k}} \right]_{z} ,
\end{displaymath}
\noindent as was generally pointed out by Haldane~\cite{Haldane}. 

\par If $\delta t_{3}$ is finite, $\Omega^{\sigma}_{\boldsymbol{k}}$ contains both odd and even components in $B$, that immediately follows from the form of $\theta^{\sigma}_{\boldsymbol{k}}$. Thus, one can write $\Omega^{\pm}_{\boldsymbol{k}} = \Omega^{o}_{\boldsymbol{k}} \pm \Omega^{e}_{\boldsymbol{k}}$, where $\Omega^{o}_{\boldsymbol{k}}$ and $\Omega^{e}_{\boldsymbol{k}}$ are, respectively, odd and even in $B$~\cite{footnote1}. This yields $\sigma_{xy} = \sigma^{{\rm I}}_{xy} + \sigma^{{\rm II}}_{xy}$ with 
\noindent
\begin{displaymath}
\sigma^{{\rm I}({\rm II})}_{xy} =  - \int_{\rm BZ} \frac{d \boldsymbol{k}}{(2 \pi)^{2}} f^{e(o)}_{\boldsymbol{k}}  \Omega^{o(e)}_{\boldsymbol{k}}, 
\end{displaymath}
\noindent where $f^{e}_{\boldsymbol{k}} = f^{+}_{\boldsymbol{k}}+f^{-}_{\boldsymbol{k}}$ and $f^{o}_{\boldsymbol{k}} = f^{+}_{\boldsymbol{k}}-f^{-}_{\boldsymbol{k}}$ are, respectively, even and odd in $B$. Thus, both $\sigma^{\rm I}_{xy}$ and $\sigma^{\rm II}_{xy}$ are odd in $B$. $\sigma^{\rm II}_{xy}$ is caused by the altermagnetic splitting of bands and is finite only when the band with one spin is occupied, while another band is empty. This is somewhat similar to ferromagnets, where AHE is caused by imperfect compensation between contributions with different spins. More intriguing is the existence of finite $\sigma^{\rm I}_{xy}$ in case of spin-degenerate bands due to special symmetry of the Hamiltonian (\ref{eq:Hk}), as was discussed above. 

\par \emph{Large-$B$ limit}. More insight can be gained by considering $\Omega^{o}_{\boldsymbol{k}}$ and $\Omega^{e}_{\boldsymbol{k}}$ in the limit of large $B$~\cite{SM}:
\noindent
\begin{displaymath}
\Omega^{o}_{\boldsymbol{k}} \approx \frac{2t_{1}t_{x}}{B|B|} \left( \frac{t_{1}^{2}-t_{x}^{2}\tan^{2}\frac{k_{x}}{2}}{t_{1}^{2}+t_{x}^{2} \tan^{2}\frac{k_{x}}{2} \tan^{2}\frac{k_{y}}{2}} \sin^{2} \frac{k_{y}}{2} - k_{x} \leftrightarrow k_{y} \right) 
\end{displaymath}
\noindent and
\noindent
\begin{displaymath}
\Omega^{e}_{\boldsymbol{k}} \approx - 2 \frac{\delta h^{0}_{\boldsymbol{k}}}{B}\Omega^{o}_{\boldsymbol{k}} +  \frac{4t_{1}t_{x} \delta t_{3}}{|B|^{3}} \left( \sin k_{x} \sin 2k_{y} - k_{x} \leftrightarrow k_{y} \right). 
\end{displaymath}
\noindent (i) The altermagnetic contribution $\Omega^{e}_{\boldsymbol{k}}$, which is proportional to $\delta t_{3}$, is expected to be smaller than $\Omega^{o}_{\boldsymbol{k}}$ by the factor $\frac{|\delta t_{3}|}{B}$. (ii) Both terms are antisymmetric with respect to the permutation of $k_{x}$ and $k_{y}$. Therefore, $\sigma_{xy}$, given by the BZ integral, will vanish unless there is some ``anisotropy'', discriminating between $k_{x}$ and $k_{y}$. The role of this anisotropy is played by the orthorhombic strain $\delta t_{2}$. The Berry curvature itself does not depend on $\delta t_{2}$. However, $\delta t_{2}$ deforms the Fermi surface [Fig.~\ref{fig:model}(e)], and thus yields finite $\sigma_{xy}$. 

\par
\begin{figure}[t]
\begin{center}
\includegraphics[width=8.6cm]{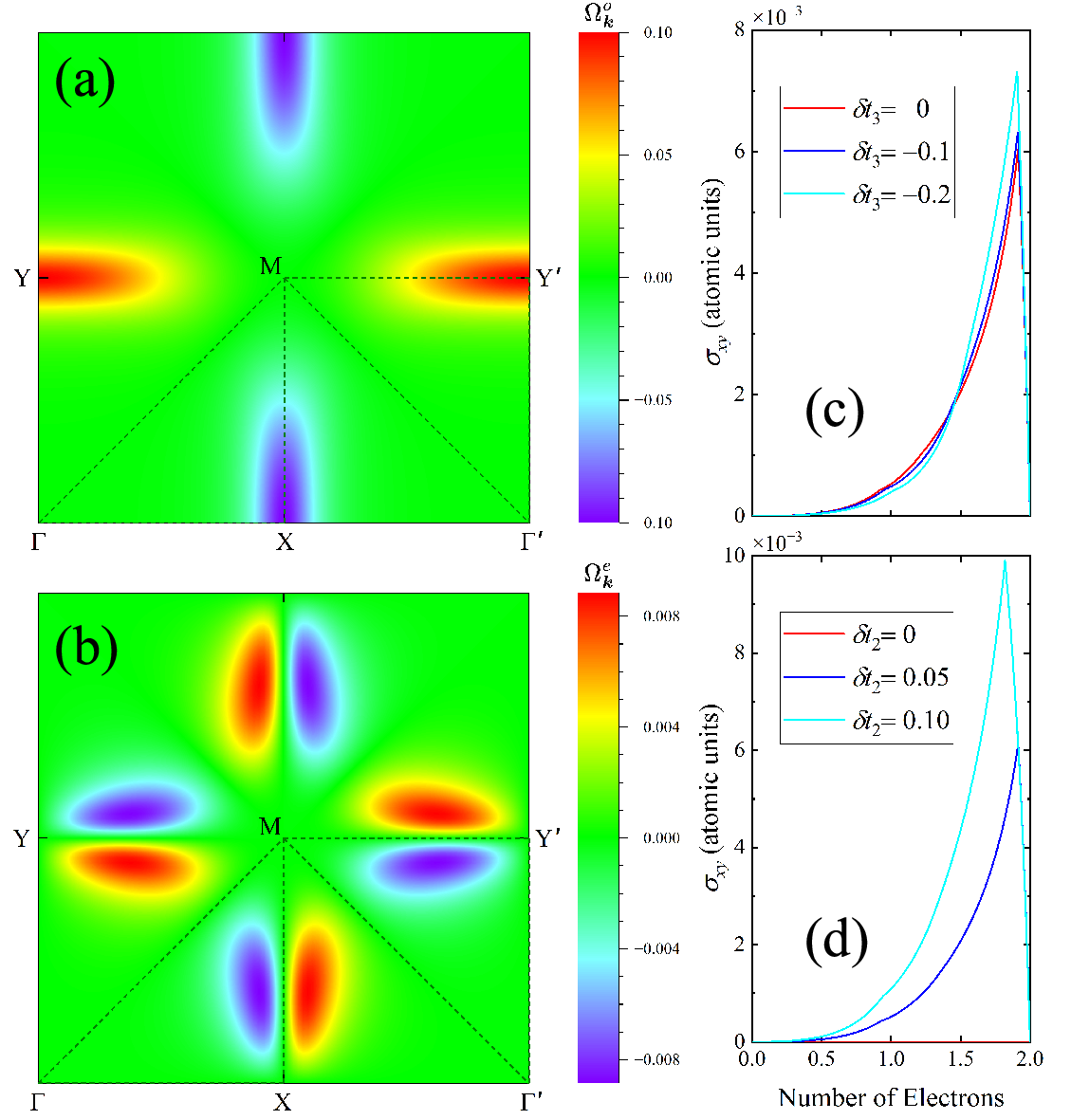} 
\end{center}
\caption{Anomalous Hall effect: (a) odd and (b) even components of $\Omega_{\boldsymbol{k}}$ in staggered magnetic field $B$; Band-filling dependence of $\sigma_{xy}$ for different values of (c) $\delta t_{3}$ and (d) $\delta t_{2}$.}
\label{fig:AHE}
\end{figure}

 The behavior of Berry curvature and $\sigma_{xy}$ is summarized in Fig.~\ref{fig:AHE}. 
\noindent
\noindent $\Omega^{o}_{\boldsymbol{k}}$ and $\Omega^{e}_{\boldsymbol{k}}$ have nodes along $\Gamma$-${\rm M}$-$\Gamma'$. Moreover, $\Omega^{e}_{\boldsymbol{k}}$ has additional nodes along ${\rm X}$-${\rm M}$-${\rm Y}'$, as can be also clearly seen from the above expressions in the large-$B$ limit. The magnitude of $\Omega^{e}_{\boldsymbol{k}}$ is generally smaller, which is again consistent with the above estimate for large $B$. The actual contribution of $\Omega^{e}_{\boldsymbol{k}}$ to $\sigma_{xy}$ is even smaller because the largest band splitting is expected along $\Gamma$-${\rm M}$-$\Gamma'$ [Fig.~\ref{fig:model}(d),(e)], which is the nodeline of $\Omega^{e}_{\boldsymbol{k}}$. Therefore, for realistic values $|\delta t_{3}| \lesssim |t_{3}|$, the contribution of $\Omega^{e}_{\boldsymbol{k}}$ to $\sigma_{xy}$ is practically negligible. Furthermore, $|\Omega^{o}_{\boldsymbol{k}}|$ is the largest along ${\rm X}$-${\rm M}$-${\rm Y}'$, where the bands are spin-degenerate. On the other hand, $\sigma_{xy}$ rises rapidly with the increase of $\delta t_{2}$. The kink of $\sigma_{xy}$ around $n_{\rm el}=1.9$ is related to the depopulation of states near the ${\rm X}$ point, which are lower in energy than the ones in the ${\rm Y}$ (${\rm Y}'$) point due to the orthorhombic strain.

\par Our toy model analysis provides a clear explanation for the behavior of AHE in AFM $\kappa$-type organic conductors and $Pbnm$ perovskites, obtained in a more sophisticated multi-orbital model, which also reveals the crucial importance of sign-alternating SOC and orthorhombic strain, and relative unimportance of the band splitting~\cite{NakaOrganic,Naka2022}. 

\par \emph{Orbital Magnetization}. ${\cal M}$ is given by the BZ integral of ${\cal M}^{\sigma}_{\boldsymbol{k}} =  {\rm Im} \langle \partial_{k_{x}} u^{\sigma}_{\boldsymbol{k}} | \hat{\cal H}^{\sigma}_{\boldsymbol{k}} + \varepsilon^{\sigma}_{\boldsymbol{k}} -2\mu  | \partial_{k_{y}} u^{\sigma}_{\boldsymbol{k}} \rangle$ (per two sites), where $\mu$ is the chemical potential~\cite{Thonhauser,Shi}. Introducing ${\cal M}^{\pm}_{\boldsymbol{k}} = {\cal M}^{o}_{\boldsymbol{k}} \pm {\cal M}^{e}_{\boldsymbol{k}}$, one can identify two contributions, ${\cal M} = {\cal M}^{\rm I} + {\cal M}^{\rm II}$, similar to AHE~\cite{SM}. In the insulating state for $n_{\rm el} = 2$, it is sufficient to consider only the orthorhombic strain part of $\hat{\cal H}^{\sigma}_{\boldsymbol{k}}$ and $\varepsilon^{\sigma}_{\boldsymbol{k}}$. Then, ${\cal M}$ is given by the BZ integral of ${\cal M}^{\rm in}_{\boldsymbol{k}} =  -2\delta h^{2}_{\boldsymbol{k}} \Omega^{o}_{\boldsymbol{k}}$.

\par
\begin{figure}[t]
\begin{center}
\includegraphics[width=8.6cm]{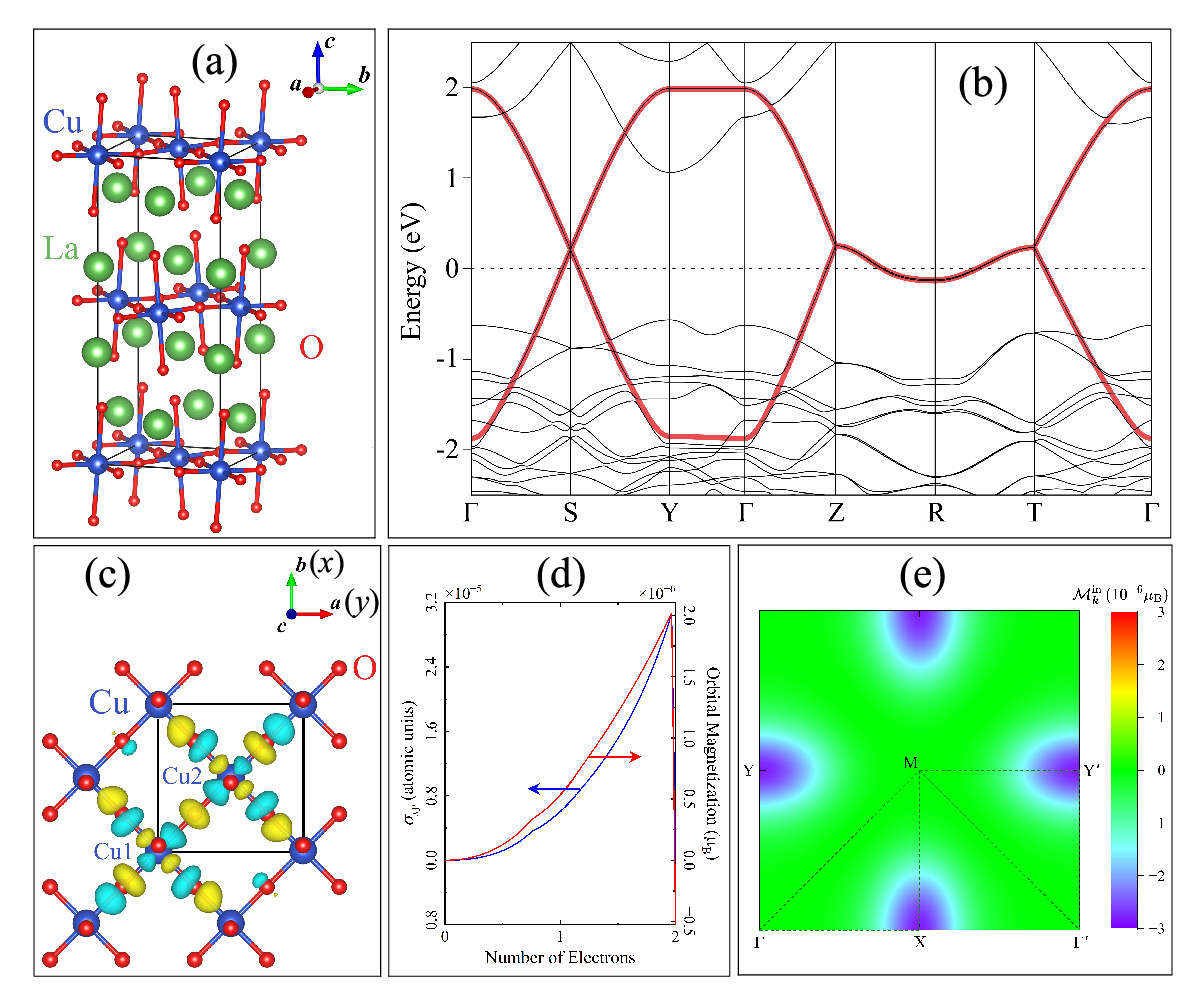} 
\end{center}
\caption{Realistic model for La$_2$CuO$_4$: (a) Crystal structure; (b) Electronic structure near the Fermi level. The red line shows the tight-binding dispersion of the $x^{2}$-$y^{2}$ band; (c) Corresponding Wannier functions; (c) Band-filling dependence of $\sigma_{xy}$ and orbital magnetization; (e) The integrand, ${\cal M}^{\rm in}_{\boldsymbol{k}}$, specifying the orbital magnetization in the insulating state for $n_{\rm el} = 2$ electrons.}
\label{fig:La2CuO4}
\end{figure}

 \emph{The case of La$_2$CuO$_4$}. The simplest realistic model for La$_2$CuO$_4$ can be constructed for the $x^{2}$-$y^{2}$ band near the Fermi level, as suggested by DFT calculations, and using for these purposes Wannier functions technique~\cite{WannierRevModPhys} (Fig.~\ref{fig:La2CuO4}). It yields (in meV)~\cite{SM}: $t_{1}=-$$439$, $t_{2}= 34$, $\delta t_{2}= 5$, $t_{3}=-$$30$, $t_{x}=-$$1$, and $t_{y}=-$$4$. $2B \approx U$ (the on-site Coulomb repulsion) can be evaluated from constrained random phase approximation as $2.2$ eV~\cite{SM}. 
\noindent
\noindent ${\cal M}$ replicates the shape of $\sigma_{xy}$, including the kink position [Fig.~\ref{fig:La2CuO4}(d)]. This is to be expected: for the narrow-band compounds, the $\boldsymbol{k}$-dispersion of $\hat{\cal H}^{\sigma}_{\boldsymbol{k}}$ and  $\varepsilon^{\sigma}_{\boldsymbol{k}}$ is relatively weak and, therefore, ${\cal M}^{\sigma}_{\boldsymbol{k}} \sim \Omega^{\sigma}_{\boldsymbol{k}}$~\cite{PRB2014}. Nevertheless, the $\boldsymbol{k}$-dispersion of $\delta h^{2}_{\boldsymbol{k}}$ additionally modulates the sign-alternating $\Omega^{o}_{\boldsymbol{k}}$ along ${\rm X}$-${\rm M}$-${\rm Y}'$, thus making ${\cal M}^{\rm in}_{\boldsymbol{k}} \le 0$ throughout the BZ [Fig.~\ref{fig:La2CuO4}(e)] and causing ${\cal M}$ to be finite. This can be viewed as a piezomagnetism induced by the orthorhombic strain. In the insulating state ${\cal M}$  is small ($\sim -5$$\times$$10^{-7} \mu_{\rm B}$), partly due to the sharp drop of ${\cal M}$ near $n_{\rm el}=2$, following a similar drop of $\sigma_{xy}$. 

\par Thus, ${\cal M}$ is considerably smaller than the spin net moment: $8St_{y}/t_{1} \sim 0.03 \mu_{\rm B}$ for $S=\frac{1}{2}$, meaning that La$_2$CuO$_4$ is the canonical weak ferromagnet. Nevertheless, this situation is not generic and there are the cases where the spin net moment vanishes, as expected from the form of DM interactions in RuO$_2$~\cite{review2024}. Then, what is the proper order parameter classifying these altermagnets~\cite{SatoHayami,OikePetersShinada}? We believe that the legitimate choice is ${\cal M}$~\cite{PRB1997}. Similar to AHE, it is ultimately related to the Berry curvature. Furthermore, although being typically small, it remains finite (unlike the spin net moment) and in many respects replicates the behavior of AHE. 

\par \emph{Summary}. Altermagnetism presents a new turn in the development of WF, bringing the analysis to the microscopic level and, thus, revealing new aspects in old-standing problems. Although from a phenomenological point of view the phenomena of WF and AHE are basically identical, the microscopic pictures behind them is different and can be linked to, respectively, nonalternating and alternating in sign DM interactions. Nevertheless, these components typically coexist as both of them are induced by the same oxygen displacements, tending to align the DM vectors perpendicular to magnetic bonds~\cite{Keffer}. This is the reason why WF and AHE are also expected to coexist. The altermagentic band splitting does not play a key role in AHE. The $\{ \mathcal{S}|{\bf t} \}$ symmetry of microscopic Hamiltonian supports the spin degeneracy of the bands, but does not exclude breaking of $\mathcal{T}$. The lack of the band splitting, which was recently observed in some potential altermagnets~\cite{MoranoMnF2}, does not necessarily mean the absence of AHE. All these features of AHE are replicated by the net orbital magnetization ${\cal M}$, which makes it the proper order parameter for classifying unconventional antiferromagnets. The orthorhombic strain, which is typically ignored in models of altermagnetism~\cite{Roig,Maier}, is the key ingredient responsible for finite AHE and ${\cal M}$ in analogy with piezomagnetism. 

\par \emph{Acknowledgement}. 
We are grateful to M. Naka, H. Seo, and A. Lichtenstein for valuable comments and discussions, M. Katsnelson for drawing our attention to the book~\cite{TurovBook}, and A. Katanin and S. Streltsov for providing a copy of this book. MANA is supported by World Premier International Research Center Initiative (WPI), MEXT, Japan.


\begin{thebibliography}{99}

\bibitem{SmejkalPRX1}
L. {\v S}mejkal, J. Sinova, and T. Jungwirth, {\em Beyond conventional ferromagnetism and antiferromagnetism: a phase with nonrelativistic spin and crystal rotation symmetry}, Phys. Rev. X \textbf{12}, 031042 (2022).

\bibitem{SmejkalPRX2}
L. {\v S}mejkal, J. Sinova, and T. Jungwirth, {\em Emergung research landscape of altermagnetism}, Phys. Rev. X \textbf{12}, 040501 (2022).

\bibitem{LingBai}
L. Bai, W. Feng, S. Liu, L. {\v S}mejkal, Y. Mokrousov, and Y. Yao, {\em Altermagnetism: exploring new frontiers in magnetism and spintronics}, Adv. Func. Patter. \textbf{34}, 2409327 (2024).

\bibitem{Naka_Spintronics}
M. Naka, Y. Motome, and H. Seo, {\em Altermagnetic perovskites}, npj Spintronics \textbf{3}, 1 (2025).

\bibitem{Noda}
Y. Noda, K. Ohno,  and S. Nakamura, {\em Momentum-dependent band spin splitting in semiconducting MnO$_2$: a density functional calculation}, Phys. Chem. Chem. Phys. \textbf{18}, 13294 (2016).

\bibitem{Okugawa}
T. Okugawa, K. Ohno, Y. Noda, and S. Nakamura, {\em Weakly spin-dependent band structures of antiferromagnetic perovskite LaMO$_3$ (M = Cr, Mn, Fe)}, J. Phys.: Condens. Matter \textbf{30}, 075502 (2018).

\bibitem{HayamiJPSJ}
S. Hayami, Y. Yanagi, and H. Kusunose, {\em Momentum-Dependent Spin Splitting by Collinear Antiferromagnetic Ordering}, J. Phys. Soc. Jpn \textbf{88}, 123702 (2019).

\bibitem{Naka}
M. Naka, S. Hayami, H. Kusunose, Y. Yanagi, Y. Motome, and H. Seo, {\em Spin current generation in organic antiferromagnets}, Nature Communications \textbf{10}, 4305 (2019).

\bibitem{SmejkalSA}
L. {\v S}mejkal, R. Gonz\'alez-Hern\'andez, T. Jungwirth, and J. Sinova, {\em Crystal time-reversal symmetry breaking and spontaneous Hall effect in collinear antiferromagnets}, Sci. Adv. \textbf{6}, eaaz8809 (2020).

\bibitem{NakaOrganic}
M. Naka, S. Hayami, H. Kusunose, Y. Yanagi, Y. Motome, and H. Seo, {\em Anomalous Hall effect in $\kappa$-type organic antiferromagnets}, Phys. Rev. B \textbf{102}, 075112 (2020).

\bibitem{Dzyaloshinskii1991}
I.~E. Dzyaloshinskii, {\em Space and time parity violation in anyonic and chiral systems}, Phys. Lett. A \textbf{155}, 62 (1991).

\bibitem{Dzyaloshinskii_weakF}
I. Dzyaloshinsky, {\em A thermodynamic theory of ``weak'' ferromagnetism of antiferromagnetics}, J. Chem. Phys. Solids \textbf{4}, 241 (1958).

\bibitem{DzyaloshinskiiPM}
I.~E. Dzyaloshinskii, {\em The problem of piezomagnetism}, Zh. Eksp. Teor. Fiz. \textbf{33}, 807 (1957) [JETP (USSR) \textbf{6}, 621 (1958)].

\bibitem{DzyaloshinskiiME}
I.~E. Dzyaloshinskii, {\em On the magneto-electrical effect in atiferromagnets}, Zh. Eksp. Teor. Fiz. \textbf{37}, 881 (1959) [JETP (USSR) \textbf{10}, 628 (1960)].

\bibitem{TurovBook}
E.~A. Turov, {\em Kinetic, optical, and acoustic properties of antiferromagnets} (Ural Division of Academy of Sciences of the USSR, Sverdlovsk, 1990).

\bibitem{TurovUFN}
E.~A. Turov, {\em Can the magnetoelectric effect coexist with weak piezomagnetism and ferromagnetism?}, Uspekhi Fizicheskikh Nauk \textbf{164}, 325 (1994) [Physics-Uspekhi \textbf{37}, 303-310 (1994)]. 

\bibitem{TurovShavrov}
E.~A. Turov and V.~G. Shavrov, {\em On some galvano- and thermomagnetic effects in antiferromagnets}, Zh. Eksp. Teor. Fiz. \textbf{43}, 2273 (1962) [JETP (USSR) \textbf{16}, 1606 (1963)].

\bibitem{Moriya_weakF}
T. Moriya, {\em Anisotropic Superexchange Interaction and Weak Ferromagnetism}, Phys. Rev. \textbf{120}, 91 (1960).

\bibitem{Katsnelson}
M.~I. Katsnelson, Y.~O. Kvashnin, V.~V. Mazurenko, and A.~I. Lichtenstein, {\em Correlated band theory of spin and orbital contributions to Dzyaloshinskii-Moriya interactions}, Phys. Rev. B \textbf{82}, 100403(R) (2010).

\bibitem{Kikuchi}
T. Kikuchi, T. Koretsune, R. Arita, and G. Tatara, {\em Dzyaloshinskii-Moriya interaction as a consequence of a doppler shift due to spin-orbit-induced intrinsic spin current}, Phys. Rev. Lett. \textbf{116}, 247201 (2016).

\bibitem{review2024}
I.~V. Solovyev, {\em Linear response theories for interatomic exchange interactions}, J. Phys.: Condens. Matter \textbf{36}, 223001 (2024).

\bibitem{PRB1997}
I.~V. Solovyev, {\em Magneto-optical effect in the weak ferromagnets LaMO$_3$ (M= Cr, Mn, and Fe)}, Phys. Rev. B \textbf{55}, 8060 (1997).

\bibitem{Thonhauser}
T. Thonhauser, D. Ceresoli, D. Vanderbilt, and R. Resta, {\em Orbital magnetization in periodic insulators}, Phys. Rev. Lett. \textbf{95}, 137205 (2005).

\bibitem{Shi}
J. Shi, G. Vignale, D. Xiao, and Q. Niu, {\em Quantum theory of orbital magnetization and its generalization to interacting systems}, Phys. Rev. Lett. \textbf{99}, 197202 (2007).

\bibitem{footnote2} In comparison with the standard setting of the $Bmab$ group, here we additionally swap the orthorhombic axes $x$ and $y$.

\bibitem{Shekhtman}
L. Shekhtman, O. Entin-Wohlman, and A. Aharony, {\em Moriya’s anisotropic superexchange interaction, frustration, and Dzyaloshinsky’s weak ferromagnetism}, Phys. Rev. Lett. \textbf{69}, 836 (1992).

\bibitem{PRL1996}
I. Solovyev, N. Hamada, and K. Terakura, {\em Crucial Role of the Lattice Distortion in the Magnetism of LaMnO$_3$}, Phys. Rev. Lett. \textbf{76}, 4825 (1996).

\bibitem{Yamaguchi}
T. Yamaguchi and K. Tsushima, {\em Magnetic symmetry of rare-earth orthochromites and orthoferrites}, Phys. Rev. B \textbf{8}, 5187 (1973).

\bibitem{footnote3}
Nevertheless, for the $Bmab$ symmetry, $\delta t_{3}$ can reappear in the multi-orbital case, considering the hoppings between orbitals belonging to different irreducible representations of the point group. 

\bibitem{Kaplan}
T.~A. Kaplan, {\em Single-Band Hubbard Model with Spin-Orbit Coupling}, Z. Phys. B \textbf{49}, 313 (1983).

\bibitem{Roig}
M. Roig, A. Kreisel, Y. Yu, B.~M. Andersen, and D.~F. Agterberg, {\em Minimal models for altermagnetism}, Phys. Rev. B \textbf{110}, 144412 (2024).

\bibitem{WannierRevModPhys}
N. Marzari, A.~A. Mostofi, J.~R. Yates, I. Souza, and D. Vanderbilt, {\em Maximally localized Wannier functions: Theory and applications}, Rev. Mod. Phys. \textbf{84}, 1419 (2012).

\bibitem{wannier}
A.~A. Mostofi, J.~R. Yates, G. Pizzi, Y.~S. Lee, I. Souza, D. Vanderbilt, and N. Marzari, {\em An updated version of Wannier90: A tool for obtaining maximally-localised Wannier functions}, Comput. Phys. Commun. \textbf{185}, 2309 (2014).

\bibitem{rpa2}
F. Aryasetiawan, M. Imada, A. Georges, G. Kotliar, S. Biermann, and A.~I. Lichtenstein. {\em Frequency-dependent local interactions and low-energy effective models from electronic structure calculations}, Phys. Rev. B \textbf{70}, 195104 (2004).
 
\bibitem{respack2}
K. Nakamura, Y. Yoshimoto, Y. Nomura, T. Tadano, M. Kawamura, T. Kosugi, K. Yoshimi, T. Misawa, and Y. Motoyama, {\em RESPACK: An ab initio tool for derivation of effective low-energy model of material}, Computer Physics Communications \textbf{261}, 107781 (2021).

\bibitem{Fang}
Z. Fang, N. Nagaosa, K.~S. Takahashi, A. Asamitsu, R. Mathieu, T. Ogasawara, H. Yamada, M. Kawasaki, Y. Tokura, and K. Terakura, {\em The anomalous Hall effect and magnetic monopoles in momentum space}, Science \textbf{302}, 92 (2003).

\bibitem{Yao}
Y. Yao, L. Kleinman, A.~H. MacDonald, J. Sinova, T. Jungwirth, D.-s. Wang, E. Wang, and Q. Niu, {\em First principles calculation of anomalous Hall conductivity in ferromagnetic bcc Fe}, Phys. Rev. Lett. \textbf{92}, 037204 (2004).

\bibitem{SM} See Supplemental Material at http://link.aps.org/supplemental/... for details of the crystal structure, construction and analysis of the model, and electronic structure calculations for La$_2$CuO$_4$, which includes Refs.~\cite{crystalstruc,pbe,qe,mp,paw,vasp}.

\bibitem{crystalstruc}
M. Reehuis, C. Ulrich, K. Proke\v{s}, A. Gozar, G. Blumberg, Seiki Komiya, Yoichi Ando, P. Pattison, and B. Keimer, {\em Crystal structure and high-field magnetism of La$_{2}$CuO$_{4}$}, Phys. Rev. B \textbf{73}, 144513 (2006).

\bibitem{pbe}
J.~P. Perdew, K. Burke, and M. Ernzerhof, {\em Generalized Gradient Approximation Made Simple}, Phys. Rev. Lett. \textbf{77}, 3865 (1996).

\bibitem{qe}
P. Giannozzi, S. Baroni, N. Bonini \textit{et al}, {\em Quantum ESPRESSO: a modular and open-source software project for quantum simulations of materials}, J. Phys.: Condens.Matter \textbf{21}, 395502 (2009).

\bibitem{mp}
H.~J. Monkhorst and J.~D. Pack, {\em Special points for Brillouin-zone integrations}, Phys. Rev. B \textbf{13}, 5188 (1976). 

\bibitem{paw}
G. Kresse and D. Joubert, {\em From ultrasoft pseudopotentials to the projector augmented-wave method}, Phys. Rev. B \textbf{59}, 1758 (1999).

\bibitem{vasp}
G. Kresse and J. Furthm\"uller, {\em Efficient iterative schemes for ab initio total-energy calculations using a plane-wave basis set}, Phys. Rev. B \textbf{54}, 11169 (1996).
    
\bibitem{Haldane}
F.~D.~M. Haldane, {\em Berry curvature on the Fermi surface: anomalous Hall effect as a topological Fermi-liquid property}, Phys. Rev. Lett. \textbf{93}, 206602 (2004).

\bibitem{footnote1}
Here, we used the property that $\Omega^{-}_{\boldsymbol{k}}$ can be obtained from $\Omega^{+}_{\boldsymbol{k}}$ by changing signs of both $B$ and $t_{x}$, which follows from the form of $\hat{\cal H}_{\boldsymbol{k}}$. 

\bibitem{Naka2022}
M. Naka, Y. Motome, and H. Seo, {\em Anomalous Hall effect in antiferromagnetic perovskites}, Phys. Rev. B \textbf{106}, 195149 (2022).

\bibitem{PRB2014}
S.~A. Nikolaev, and I.~V. Solovyev, {\em Orbital magnetization of insulating perovskite transition-metal oxides with a net ferromagnetic moment in the ground state}, Phys. Rev. B \textbf{89}, 064428 (2014).

\bibitem{SatoHayami}
T. Sato and S. Hayami, {\em Quantum theory of magnetic octupole in periodic crystals and characterization of time-reversal-symmetry breaking antiferromagnetism}, arXiv:2504.21431 (2025).

\bibitem{OikePetersShinada}
J. Oike, R. Peters, and K. Shinada, {\em Thermodynamic formulation of the spin magnetic octupole moment in bulk crystals}, arXiv:2504.21418 (2025).

\bibitem{Keffer}
F. Keffer, {\em Moriya Interaction and the Problem of the Spin Arrangements in $\beta$MnS}, Phys. Rev. \textbf{126}, 896 (1962).

\bibitem{MoranoMnF2}
V.~C. Morano, Z. Maesen, S.~E. Nikitin, J. Lass, D.~G. Mazzone, and O. Zaharko, {\em Absence of altermagnetic magnon band splitting in MnF$_2$}, Phys. Rev. Lett. \textbf{134}, 226702 (2025).

\bibitem{Maier}
T.~A. Maier and S. Okamoto, {\em Weak-coupling theory of neutron scattering as a probe of altermagnetism}, Phys. Rev. B \textbf{108}, L100402 (2023).

\end{thebibliography}
\end{document}